\documentclass[%
 reprint,
superscriptaddress,
showpacs,
amsmath,amssymb,
prb,
floatfix,
]{revtex4-1}
\usepackage{graphicx}
\usepackage{subfigure} 
\usepackage[colorlinks,linkcolor=blue,anchorcolor=green,citecolor=blue]{hyperref}
\usepackage{bm}%

\begin{document}

\title{State transition and electrocaloric effect of BaZr$_{x}$Ti$_{1-x}$O$_3$: simulation and experiment}

\author{Yang-Bin Ma}
 \email{y.ma@mfm.tu-darmstadt.de}
\affiliation{Institute of Materials Science, Technical University of Darmstadt, 64287 Darmstadt Germany}

\author{Christian Molin}
\affiliation{Fraunhofer Institute for Ceramic Technologies and Systems, 01277 Dresden Germany}

\author{Vladimir V. Shvartsman}
\affiliation{University of Duisburg-Essen and Center for
Nanointegration Duisburg-Essen (CENIDE), 45141 Essen Germany}

\author{Sylvia Gebhardt}
\affiliation{Fraunhofer Institute for Ceramic Technologies and Systems, 01277 Dresden Germany}

\author{Doru C. Lupascu}
\affiliation{University of Duisburg-Essen and Center for
Nanointegration Duisburg-Essen (CENIDE), 45141 Essen Germany}

\author{Karsten Albe}
\affiliation{Institute of Materials Science, Technical University of Darmstadt, 64287 Darmstadt Germany}

\author{Bai-Xiang Xu}
 \email{xu@mfm.tu-darmstadt.de}
\affiliation{Institute of Materials Science, Technical University of Darmstadt, 64287 Darmstadt Germany}

\date{\today}

\begin{abstract}

The electrocaloric effect (ECE) of BaZr$_{x}$Ti$_{1-x}$O$_3$ (BZT) is closely related to the relaxor state transition of the materials. This work presents a systematic study on the ECE and the state transition of the BZT, using a combined canonical and microcanonical Monte Carlo simulations based a lattice-based on a Ginzburg-Landau-type Hamiltonian. 
For comparison and verification, experimental measurements have been carried on BTO and BZT ($x=0.12$ and $0.2$) samples, including the ECE at various temperatures, domain patterns by Piezoresponse Force Microscopy at room temperature, and the P-E loops at various temperatures. 
Results show that the dependency of BZT behavior of the Zr-concentration can be classified into three different stages. 
In the composition range of $ 0 \leq x \leq 0.2 $, ferroelectric domains are visible, but ECE peak drops with increasing Zr-concentration harshly. 
In the range of $ 0.3 \leq x \leq 0.7 $, relaxor features become prominent, and the decrease of ECE with Zr-concentration is moderate. In the high concentration range of $ x \geq 0.8 $, the material is almost nonpolar, and there is no ECE peak visible. 
Results suggest that BZT with certain low range of Zr-concentration around $x=0.12 \sim 0.3 $ can be a good candidate with relatively high ECE and simutaneously wide temperature application range at rather low temperature.

\end{abstract}

\maketitle


\section{Introduction}\label{sec:intro}


The electrocaloric effect (ECE) has been shown to be attractive for solid refrigeration application.~\cite{2012_Fahler}
As one of the most simple relaxors and environment-friendly ECE materials, the solid solution system BaZr$_{x}$Ti$_{1-x}$O$_3$ (BZT) has attracted great interest. 
Qian \textit{et al.}~\cite{2014_Qian} measured the ECE of BaZr$_{0.2}$Ti$_{0.8}$O$_3$ (BZT-20) samples and reported a large temperature variation of 4.5\,K under 14.5\,kV\,mm$^{-1}$ with a 30\,K operating temperature range, which is wider than that in BaTiO$_3$ (BTO). Ye \textit{et al.}~\cite{2014_Ye} studied the same kind of material and published an even higher temperature variation of $\Delta T=7$\,K under 19.5\,kV\,mm$^{-1}$.
These observations suggest that BZT might be a promising candidate for the electrocaloric cooling.

However, the mechanism of the ECE in BZT is still not fully understood. 
In fact, the origin of the relaxor behavior and state transition in BZT, to which the ECE of BZT is closely related, is still under debate. 
There are different opinions and models on the relaxor behavior, e.g., the concept of polar nanoregions~\cite{2006_Bokov}, the dipolar-glass model~\cite{1954_Smolenskii}, the random field
model~\cite{1992_Westphal,1996_Glinchuk}, the spherical random bond-random field model~\cite{1999_Pirc} and the like.
There have been a number of experimental studies on BZT to reveal its relaxor behavior from various aspects.~\cite{2002_Yu,2006_Bokov,2006_Maitib,2006_Dixit,2007_Liang,2008_Ke,2007_Xiong,2009_Karan,2008_Maiti,2010_Ricinschi,2012_Shvartsman,
2011_Maiti,2014_Ye,2015_Kleemann} 
It is believed that the small size-difference between Ti$^{4+}$ and Zr$^{4+}$ ions induces random stain fields, which are much weaker than other relaxors with heterovalent cation substitution.~\cite{2006_Dixit,2009_Shvartsman,2011_Maiti,2015_Kleemann} 
Hence such kind of random field theory may not be appropriate to explain the relaxor behavior of BZT.
The long-range Ti-O-Ti-O bonds that give rise to the dipolar correlation are broken through the substitution of Ti$^{4+}$ by non-off-center Zr$^{4+}$.~\cite{2015_Kleemann}
This mechanism was used to explain the domain patterns in BZT. 
On the other hand, the first principles calculations were performed to study different properties of BZT.~\cite{2009_Bilic,2010_Laulhe,2013_Sherrington} 
Using a lattice-based model Padurariu et al.~\cite{2011_Padurariu} simulated the relaxor behavior of BaM$_x$Ti$_{1-x}$O$_3$ (M=Zr, Sn, Hf).
They incorporated permanent dipoles caused by Ti ions, O-related induced dipoles and zero dipoles caused
by M ions to calculate the local field.

In the current work, we present Lattice-based Monte Carlo simulation results by using a Ginzburg-Landau type effective Hamiltonian which reveal the ECE and the state transition of BZT with various Zr concentration.
In the authors' previous work~\cite{2015_Maa}, a generic model based on Ginzburg-Landau type effective Hamiltonian was proposed for relaxor ferroelectrics, which was applied to study the influence of random fields on the relaxor behavior.
In the the current paper, a more dedicated effective Hamiltonian is proposed for BZT. In particular, different ground state Landau-type terms are introduced for the sites occupied by Ti-located unit cells and by Zr-located unit cells. A multi-well Landau term is used for the polar Ti-located unit cells, while a single-well Landau term is applied for the nonpolar/weakly polar Zr-located unit cells. Moreover, the dipole-dipole interaction term becomes also dependent on the nature of the two involved dipoles. This aspect is considered in the model by adjusting the high-frequency permittivity parameters in the dipole-dipole energy.  
As it has been shown in our work~\cite{2015_Maa}, one advantage of the Monte Carlo simulations is that the combined canonical and microcanonical ensemble allows us to evaluate the ECE directly without using the Maxwell relation. 
The technical details for the experiments are elaborated in Subsec.~\ref{subsec:expe}. 
The details about the model and the simulation setups can be found in Subsec.~\ref{subsec:model}, and the utilized algorithms are explained in App.~\ref{sec:alg}.

The experimental and simulation results on the ECE of BZT is presented in Sec.~\ref{sec:ece}. The influence of Zr content on ECE is discussed in details. Since the ECE of BZT is correlated to the transition state of BZT, thereafter, the domain pattern and P-E loops of ceramics samples of BZT with 0\%, 12\% and 20\% Zr content were measured experimentally, and compared with the simulations results . 
Together with the results on more compositions, the ferroelectric to relaxor ferroelectric state transition of BZT is disclosed through the simulation results on the temperature-induced polarization variation, the Zr-concentration dependency of the domain structure and the P-E loops in Sec.~\ref{sec:eq}.

\section{Details on the experiment and model}

In Subsec.~\ref{subsec:expe} and Subsec.~\ref{subsec:model} the experimental techniques and the simulation details are explained, respectively.

\subsection{Experiment details}\label{subsec:expe}

\begin{figure}[h]
  \centering
  \centerline{\includegraphics[width=8cm]{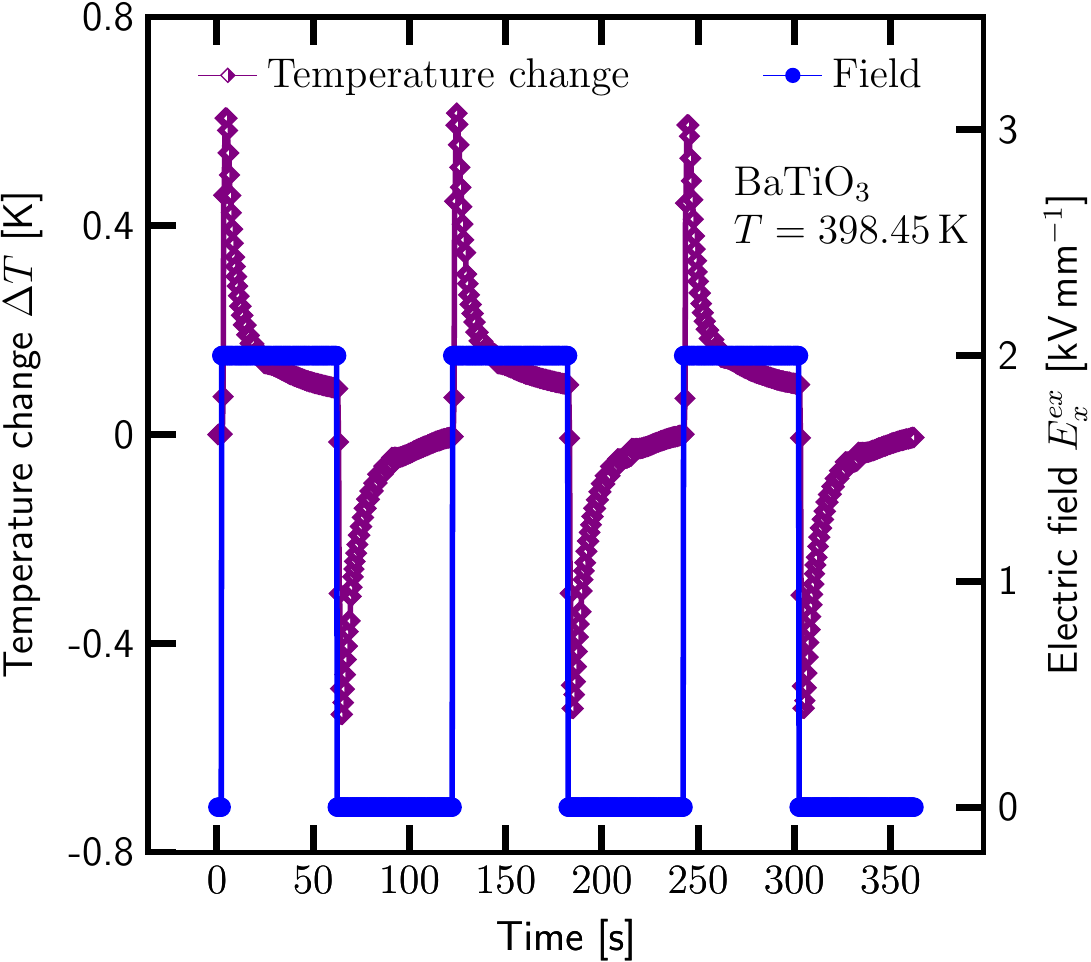}}
   \caption{An example measurement of BaTiO$_3$ around 398.15\,K. The electric field of 2 kV/mm was applied suddenly. After application, the electric field was switched on for 60\,s and then switched off for another 60\,s.}
 \label{fig:figure1}
\end{figure} 
%
Ceramics samples of BZT with 0\%, 12\% and 20\% Zr content were synthesized using a mixed oxide route.
Raw materials, BaCO$_3$ (99.86\%, Solvay, Italy), ZrO$_2$ ($\geq$ 99\%, Saint-Gobain, France) and TiO$_2$ (99.5\%, Tronox TR-HP2, Germany) were mixed in the stoichiometric ratio and milled in water using a planetary ball mill (Fritsch, Pulverisette 5, Germany) at 200 rpm for 6 hours.

%

The powders were dried, sieved and calcined in an alumina crucible at 1473.15\,K for 2\,h.
The calcined powders were subsequently pressed into disc samples with 10\,mm in diameter and sintered at 1598.15\,K for 4\,h with a heating rate of 5\,K/min.
All sintered discs have a diameter around 8.5\,mm and a thickness of about 1\,mm.
The porosity of the samples was determined by area analysis using Axiovision program (Rel. 4.7.2.0, Zeiss AG, Germany).
Respectively, the porosity is 1.3$\pm$0.8\% for BTO, 2.1$\pm$0.3\% for BaZr$_{0.2}$Ti$_{0.8}$O$_3$ (BZT-12) and 3.1$\pm$0.6\% for BZT-20.

In the ECE measurement sheath thermocouples are utilized. The detailed loading history is explained in Fig.~\ref{fig:figure1}. For hysteresis measurements, electrodes of silver conductive pastes were painted and fired at 1073.15\,K for 20\,min.
The hysteresis loop measurements were carried out at 5\,Hz, using a Sawyer Tower circuit with an analog input and digital I/O module (Measurement Computing, USB-1616HS-4).

\subsection{Model}\label{subsec:model}

For the lattice-based Monte Carlo simulations, a Ginzburg-Landau type effective Hamiltonian is used, similar as in our previous work~\cite{2015_Maa,2015_Mab,2016_Ma}. 
The potential Hamiltonian $E$ includes four contributions:
the static ground state energy $E_D$, the dipole-dipole interaction energy
$E_{dip}$, the gradient energy $E_{gr}$, and the electrostatic energy $E_e$.
\begin{equation}
E = E_D + E_{dip} + E_{gr} + E_e 
\end{equation}

The static ground state energy $E_D$ can be expressed as in the 2D case:
\begin{eqnarray} \label{eq:landau}
 {E_D}=&&V_0\sum_{i}\biggr[\dfrac{a_i}{2}(P_x^2(\mathbf{r}_i)+P_y^2(\mathbf{r}
_i)) \nonumber \\
+&&\dfrac{b_i}{4} (P_x^4(\mathbf{r}_i)+P_y^4(\mathbf{r}_i))+
    \dfrac{c_i}{6}(P_x^6(\mathbf{r}_i)+P_y^6(\mathbf{r}_i))\biggr],
\end{eqnarray}
where $V_0=a_0^3$ is the volume per site with $a_0$ being the lattice constant, $a_i,b_i,c_i$ are material-dependent coefficients, $\mathbf{r}_i$ is the coordinate of site $i$, and 
$P_x(\mathbf{r}_i)$, $P_y(\mathbf{r}_i)$
are the Cartesian components of the polarization vector
$\mathbf{P}(\mathbf{r}_i)$ at site $i$. It is assumed that the lattice constant $a_0=4$~\AA ~is independent of the Zr content, since the lattice constants change insignificantly with Zr substitution.

\begin{figure}[htp]
  \centering
  \centerline{\includegraphics[width=7cm]{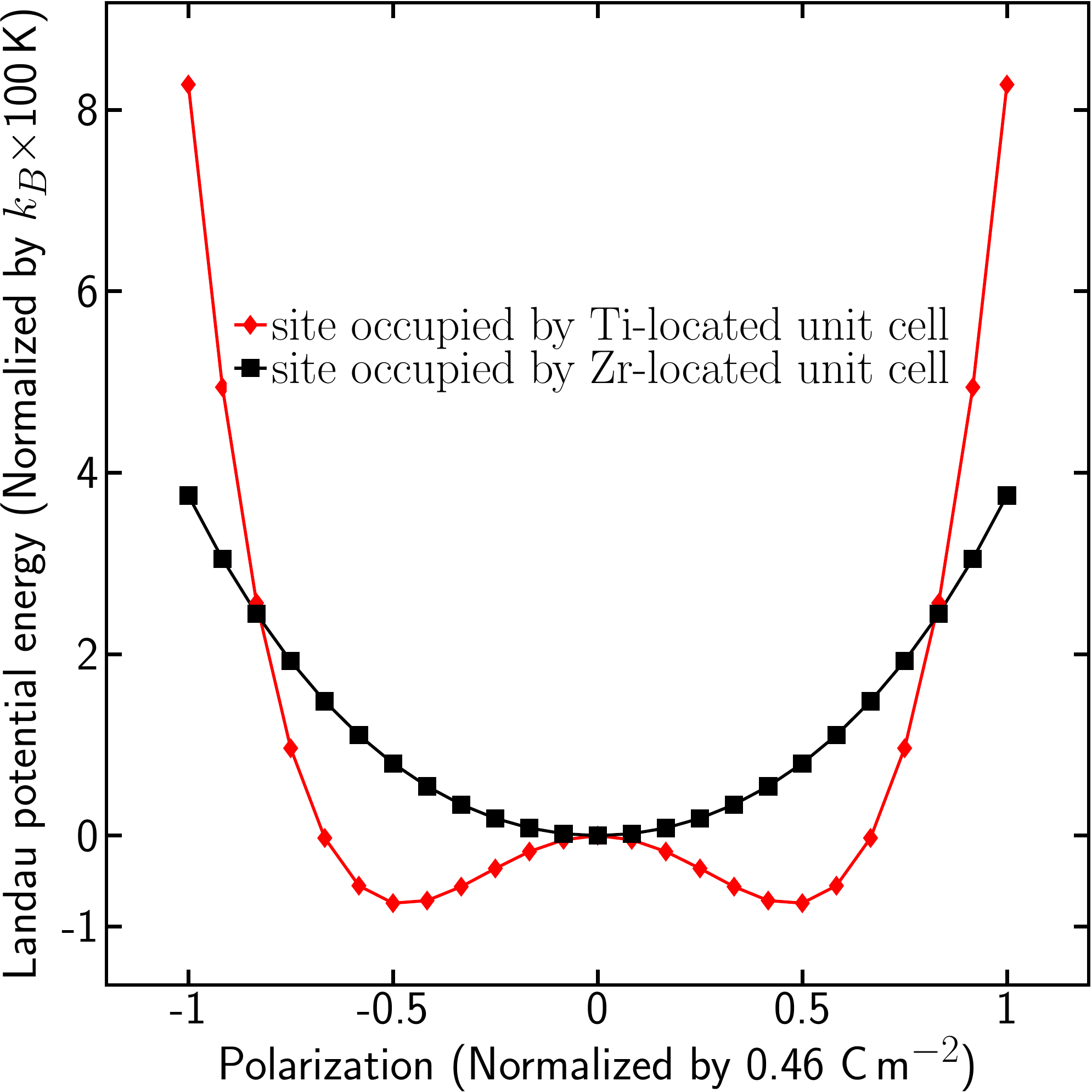}}
   \caption{A multi-well Landau-type term is applied for the sites occupied by Ti-located unit cells, and a single well type for Zr-located unit cells. }
 \label{fig:figure2}
\end{figure}

BTO is a ferroelectric material while BaZrO$_3$ is paraelectric.
Therefore, for the sites occupied by Ti-located unit cells $E_D$ is interpreted as a Ginzburg-Landau multi-well energy term while a 
single-well energy term for the site occupied by Zr-located unit cells. This concept for simulation is illustrated in Fig.~\ref{fig:figure2}. 
Based on first-principles results with zero strain and no non-soft-mode eigenmode amplitudes~\cite{1994_King-Smith} the necessary coefficients can be obtained, which are shown in the following with the sixth order term ignored: 

\begin{widetext}
\begin{equation} \label{eq:landauparameter}
 \begin{cases}
  a_i=-13.7128 \times 10^8 \text{\,J\,m\,C}^{-2}, 
  b_i=28.908\times 10^9 \text{\,J\,m}^5 \text{\,C}^{-4} 
  & \text{for Ti-located unit cells},
 \\
  a_i=6.112 \times 10^8 \text{\,J\,m\,C}^{-2}, 
  b_i=1.4454\times 10^9 \text{\,J\,m}^5 \text{\,C}^{-4} 
 & \text{for Zr-located unit cells}.
 \end{cases}
\end{equation}
\end{widetext}

The dipole-dipole interaction energy~\cite{2008_Rabe} has the expression
\begin{eqnarray} \label{eq:dipole_inter}
{E_{dip}}&=& V_0^2 \dfrac{1}{8\pi\epsilon_0 } \sum_{i,j}
\dfrac{1}{\epsilon _{ij}}
\biggr[
\dfrac{\mathbf{P}(\mathbf{r}_i) \cdot
\mathbf{P}(\mathbf{r}_j)}{|\mathbf{r}_i-\mathbf{r}_j|^3}\nonumber\\
&-&\dfrac{3[\mathbf{P}(\mathbf{r}_i) \cdot (\mathbf{r}_i-\mathbf{r}_j)]
[\mathbf{P}(\mathbf{r}_j) \cdot (\mathbf{r}_i-\mathbf{r}_j)]
}{|\mathbf{r}_i-\mathbf{r}_j|^5} \biggr],
\end{eqnarray}
where $\epsilon_0$ is the vacuum permittivity.
$\epsilon_{ij}$ is the high-frequency permittivity, which depends on Zr concentration.
Different values of the high-frequency permittivity $\epsilon_r$ for BTO were given in literature, varying from 6.0 to 15.0.~\cite{1980_Servoin,2006_Hlinka,2014_Nayak}
In the present paper, we choose $\epsilon_r=12.0$ for the case when both sides $i$ and $j$ are occupied by Ti-located unit cells. 

The permittivity of BaZrO$_3$ is chosen to be 144. Note that the value of high-frequency permittivity for BaZrO$_3$ reported in the literature~\cite{2005_Akbarzadeh} based on the first-principles calculation is close to the value of BTO. However, we argue for the choice of higher permittivity for BaZrO$_3$ as follows. 
Firstly, in their first-principles calculations the short-range coefficient $j_5$, which reflects the antiferroelectric behavior~\cite{2012_Akbarzadeh}, is much smaller in BaZrO$_3$ than in BTO.
Since in the present MC model, the short range interaction coefficient $j_5$ shown in first-principles study is not included for the simplification, the high frequency permittivity should be modified. 
Moreover, experiments~\cite{2002_Pantou,2008_Maiti,2009_Shvartsman} showed that the real part of the permittivity is smaller when increasing Zr content in BZT. It is theoretically known that the decrease of high frequency permittivity value leads to the increase of the real part of the permittivity. Thus the experimental results~\cite{2002_Pantou,2008_Maiti,2009_Shvartsman} indicate that higher permittivity value should be used for BZT than for BTO. 
To ensure BaZrO$_3$ to be a paraelectric material, the high frequency permittivity of BZT should be high enough.
Since no additional data for the high permittivity of BZT is available in the literature, the composition-dependent permittivity of BZT is simply approximated by the linear interpolation of the two ends, i.e., pure BTO and BaZrO$_3$. 
In a short summary, $\epsilon_{ij}$ is taken as
\begin{widetext} 
\begin{equation} \label{eq:dip-par}
{\epsilon _{ij}} = 
 \begin{cases}
  12.0 
  & \text{if both sites } i \text{ and } j \text{ are occupied by Ti-located unit cells},
 \\
  144.0 
  & \text{if both sites } i \text{ and } j \text{ are occupied by Zr-located unit cells},
 \\
  12.0(1-x) + 144.0 x 
  & \text{if sites } i \text{ and } j \text{
 are occupied by different unit cells},
 \end{cases}
\end{equation}
\end{widetext}
where $x$ is the Zr concentration in BZT.

It should be noted that this concept of using the Zr-concentration-dependent high-frequency permittivity should be similar to the random-bond concept proposed by Pirc and Blinc~\cite{1999_Pirc}.

We mimic the domain wall energy~\cite{1990_Cao,2004_Liub} by the gradient energy:
\begin{eqnarray} \label{eq:gradient}
  {E_{gr}} &=& S_J V_0 \sum_{i} \biggr[ g_1 ( P_{x,x}^2(\mathbf{r}_i) +
P_{y,y}^2(\mathbf{r}_i))                   \nonumber
  \\
 &+&
  g'_{1}P_{x,x}(\mathbf{r}_i) P_{y,y} (\mathbf{r}_i)    \nonumber
 + g_{2} ( P_{x,y}(\mathbf{r}_i) + P_{y,x} (\mathbf{r}_i) )^2 \nonumber
  \\
 &+&g_{2}^\prime
(P_{x,y} (\mathbf{r}_i) -
  P_{y,x} (\mathbf{r}_i) ) ^2 \biggr],
  \end{eqnarray}
where the comma in the subscripts denote derivatives with respect to the following coordinate, $S_J$ is a gauge coefficient, and the constants $g_1$, $g'_1$, $g_2$, $g_2^\prime$ allow anisotropic contribution of the gradient term. The parameters for the domain wall energy is set to be the same as these for BTO in our previous work~\cite{2015_Maa}.
On the other hand, the size-difference between Ti$^{4+}$ and Zr$^{4+}$ is small, and thus the related elastic energy and the electrostrictive interaction should be negligible. This simplification should not have qualitative influence on the calculated entropy variations.

The electrostatic energy $E_e$ is given as:
\begin{equation} \label{eq:electrostatic}
 E_e=-V_0 \sum_{i} \biggr[ \mathbf{P}(\mathbf{r}_i) \cdot
\mathbf{E}^{ex} \biggr],
\end{equation}
where $\mathbf{E}^{ex}$ is the external electric field.

%

%


\section{Electrocaloric effect}\label{sec:ece}

\begin{figure}[htp]
  \centering
  \centerline{\includegraphics[width=8cm]{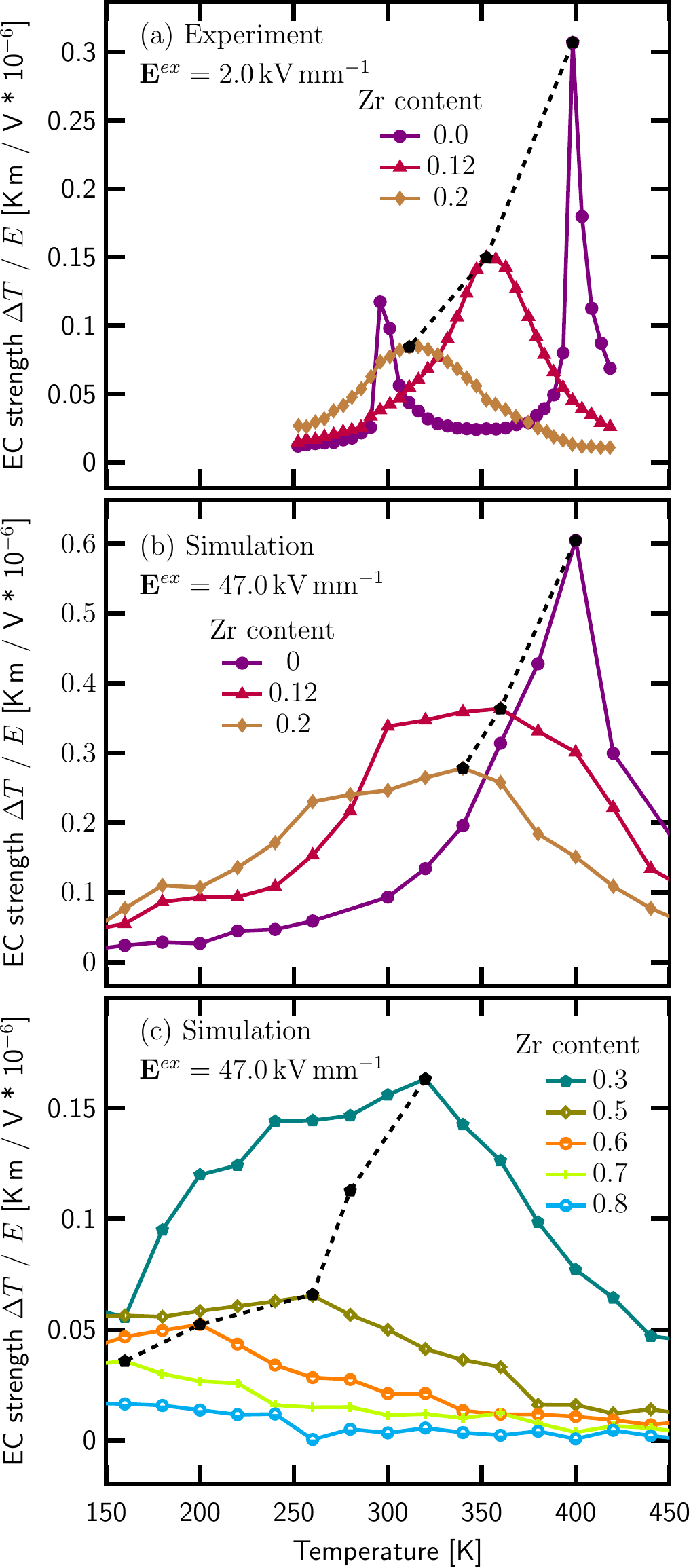}}
   \caption{
   (a) the electrocaloric effect measurement , and (b)-(c) simulation.
   In the experiment (a) and simulation (b), the temperature and Zr-content dependence of the electrocaloric effect was revealed for BaZr$_{x}$Ti$_{1-x}$O$_3$ with $0\leq x \leq 0.2$. 
   When the Zr content is slightly increased, a sharp decrease of the electrocaloric effect can be observed. With respect to this point, the experimental results agree with the simulated conclusion qualitatively. 
   In the simulation (c), the temperature and Zr-content dependence of the electrocaloric effect was revealed for BaZr$_{x}$Ti$_{1-x}$O$_3$ with $ x \geq 0.3$. 
   Two phases of ECE can be distinguished: a moderate decrease with $ 0.3 \leq x \leq0.7 $, and no peak existence with $x \geq 0.8 $. 
   The ECE peaks both in the simulation and experiment are hooked up by dashed lines with black squares.}
 \label{fig:figure3}
\end{figure} 
In this section, the ECE of BZT is experimentally measured and theoretically calculated using the direct approach. Particularly, the influence of Zr concentration is elaborated. 

Prepoled samples through MC canonical assemble are applied with an external field of $E^{ex}= 47.0$kV mm$^{-1}$ through MC microcanonical asssemble, in order to evaluate the ECE under adiabatic condition. More details about the loading history and the algorithms can be found in the previous paper~\cite{2015_Maa}. 

%

\begin{figure*}[htp]
  \centering
  \centerline{\includegraphics[width=17cm]{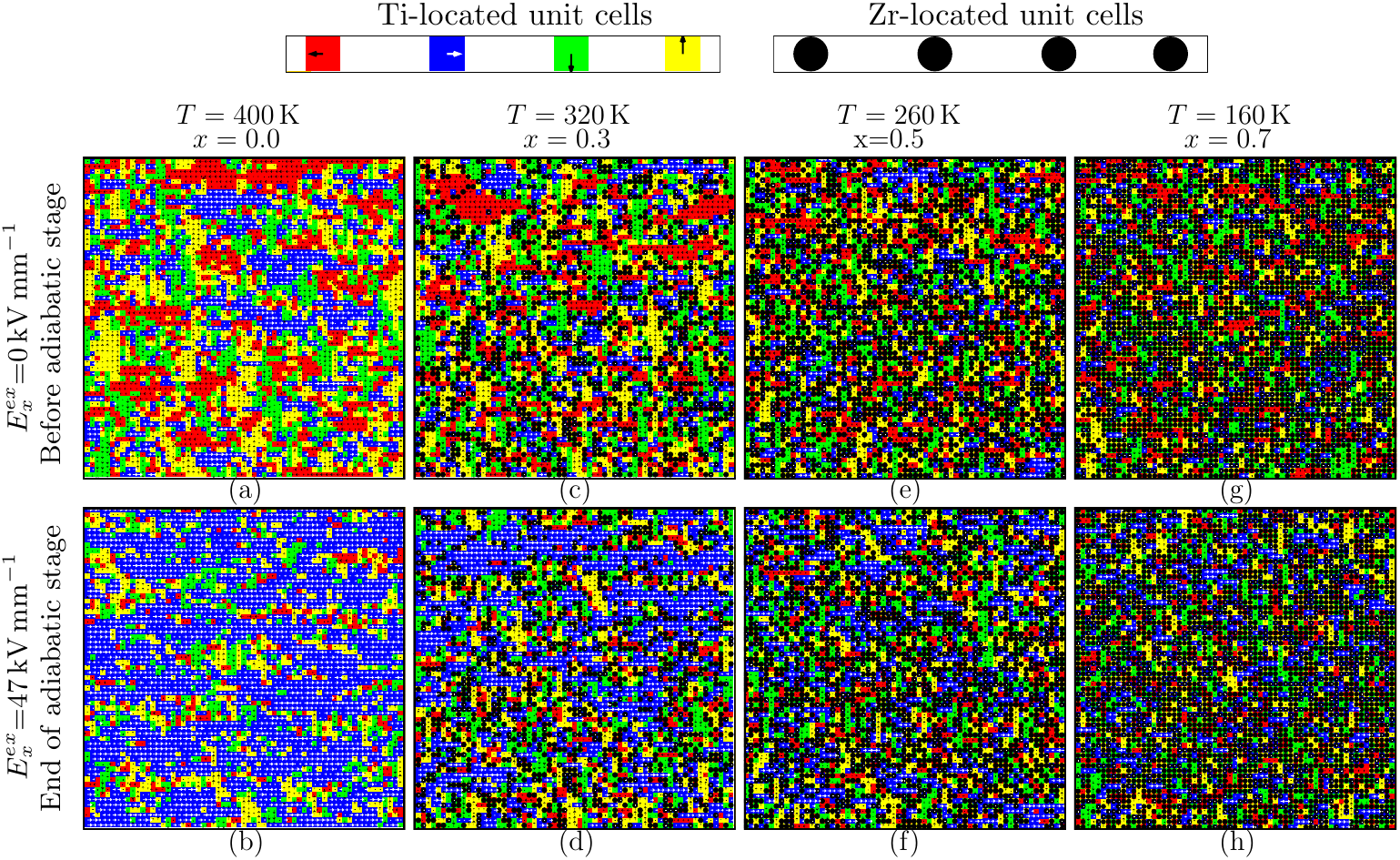}}
   \caption{Explanation to the ECE influence of the Zr content in BaZr$_{x}$Ti$_{1-x}$O$_3$ by domain pattern in the simulation. With $0 \leq x \leq 0.2$, the number of sites responsible for the configurational space $N_{effective}$ increases sharply at the end of the adiabatic stage. This signifies the sharp drop of ECE peak within this range. With $ 0.3 \leq x \leq 0.7 $, $N_{effective}$ becomes saturating gradually upon increasing Zr content. Therefore, the value of ECE peak decreases more and more moderately.    
   The red, blue, yellow and green squares
   represent the dipoles pointing respectively to the left, right, top and bottom. The black dots represent the sites occupied by Zr-located unit cells. }
 \label{fig:figure4}
\end{figure*}

\begin{figure}[htp]
  \centering
  \centerline{\includegraphics[width=7cm]{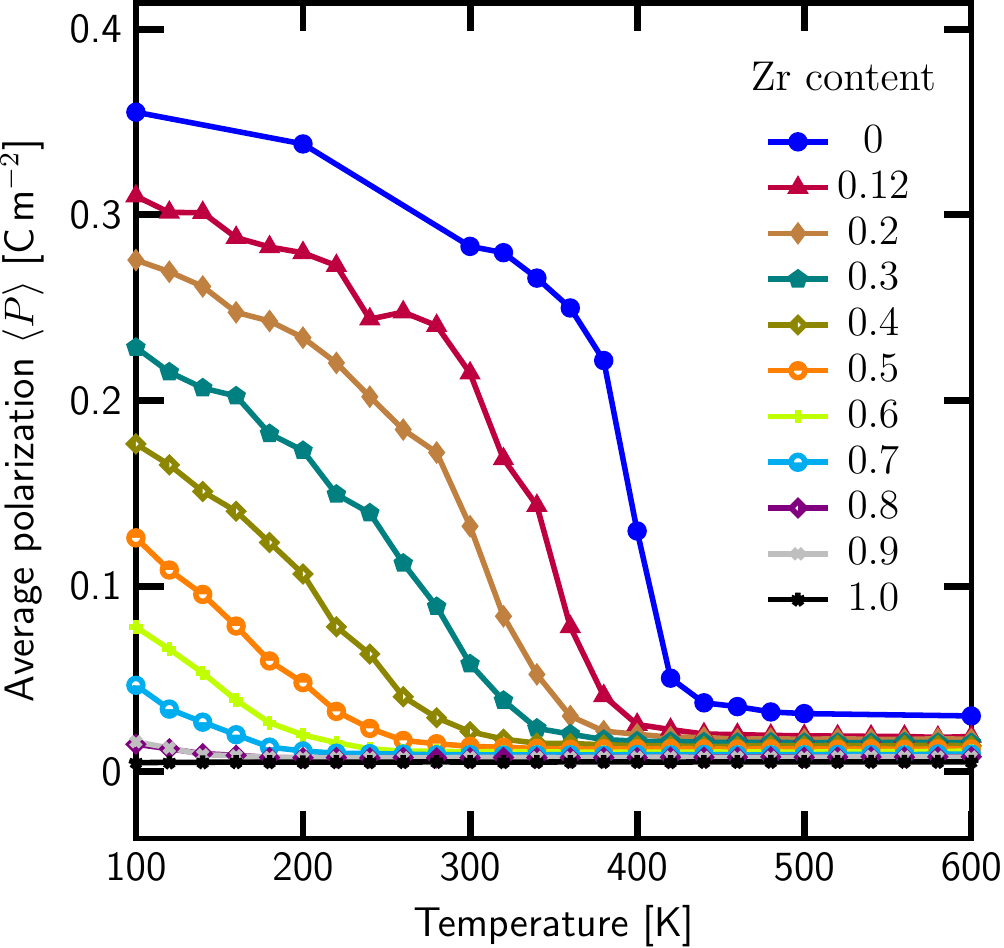}}
   \caption{The temperature-induced polarization for different Zr concentrations of BaZr$_{x}$Ti$_{1-x}$O$_3$ in the simulation. 
   The legend presents the Zr content $x$. 
   Since more sites are occupied by Zr-located unit cells, i.e., the weakly polar unit cells, are introduced, the change of the polarization becomes more moderate with respect to temperature, and the polarization becomes fairly small even at low temperature. }
 \label{fig:figure5}
\end{figure}

The Zr-content dependence of the ECE at different temperatures is demonstrated in Fig.~\ref{fig:figure3}. 
The peak value of temperature change for each composition is marked as black squares.
In general, as Zr concentration increases, the peak shifts to lower temperature. 
ECE peaks appear at the temperature where there is a sharp change of the state of the order. The state transition can be identified in Fig.~\ref{fig:figure5}. 
When Zr concentration is higher, less thermal energy are required to achieve the change from the order to disorder, i.e., the point at which the state of the order varies shifts to a lower temperature.

At first, the experiment was carried out to measure the ECE in BTO, BZT-12 and BZT-20, using sheath thermocouples. 
The measured ECE strength $\Delta T / \Delta E$ is plotted in Fig.~\ref{fig:figure3}(a), and the ECE peaks are marked using black squares.
The detailed loading history was already explained in Fig.~\ref{fig:figure1}. 
At temperature where the tetragonal phase transforms to cubic phase in BTO (398.45\,K), BZT-12 (352.45\,K) and BZT-20 (316.19\,K) or the orthorhombic phase transforms to tetragonal phase in BTO (295.9\,K), the ECE peaks appear. 
Figure~\ref{fig:figure3}(a) further demonstrates that with increasing Zr-content the ECE peak drops quickly, and the ECE peak shifts to lower temperature. 
On the other hand, the application temperature range becomes wider with more Zr substitution.

Moreover, BTO, BZT-12 and BZT-20 are theoretically investigated in Fig.~\ref{fig:figure3}(b).
It is noticeable that the peak of the temperature variation drops harshly. This significant change can be interpreted by the corresponding domain patterns. The domain pattern at each ECE peak was shown in Fig.~\ref{fig:figure4}.
In the initial state (the prepoled sample), the configurational entropy $S_c$ is apparently higher in BTO than in BZT-12. 
Nonetheless, at the end of the adiabatic stage, similar domain patterns and comparable $S_c$ are revealed for both BTO and BZT-12. 
In short, the variation of $S_c$, i.e., the variation of the temperature, in BTO is bigger than in BZT-12. 
The sites occupied by Zr-located unit cells are energetically unstable sites, and their neighbor sites become more fluctuating since mutual short- and long-interactions are present. 
Since the long-range interaction is present, with $0.1 \leq x \leq 0.2$ the effective number of the sites responsible for the configurational space $N_{effective}$ increases much more than linearly 
with the increasing of Zr-concentration, which signifies the sharp declination of the temperature variation.
In short, with increasing the Zr content, the ECE peak is lowered and appears at lower temperature. These two phenomena have been captured both in our experimental and simulated results.

Despite of the good qualitative agreements, there are notable discrepancies between the measured and calculated temperature changes and the applied electric field. These discrepancies can be explained as follows. 
Firstly, compared with the experiment, only one ECE peak can be observed in the simulation results for pure BTO in Fig.~\ref{fig:figure3}(a). It is due to the fact that in the simulation only the transition from the tetragonal to the cubic phase is considered. 
Secondly, the samples used in the simulation are supposed without defects and are single crystalline thin film materials while the samples in the measurements are ceramics. Hence, the applied external field and the ECE in the simulation is much higher than in the experiments. If compared with the experimental value of thin film BZT-20 (7\,K temperature variation under 19.5\,kV\,mm$^{-1}$)~\cite{2014_Ye}, the discrepancy would be much less.
Thirdly, the 2D simplification in the simulations underestimates the heat capacity, and thus the temperature variation in 2D should be bigger than in 3D.
Fourthly, at the ECE peak of BTO, the calculated $\Delta T / \Delta E$ is equal to $0.59 \times 10 ^{-6} $\,K $\cdot$ \,V$^{-1}$\,m, which is 
1.8 times of the experimental data $0.32\times 10 ^{-6} $\,K $\cdot$ \,V$^{-1}$\,m. 
Since our ECE measurement in the experiments is not strictly under adiabatic condition, which reduces the temperature variation value, if compared with the experimental data $0.83\times 10 ^{-6} $\,K $\cdot$ \,V$^{-1}$\,m by Moya~\cite{2014_Moya}. 
The measurement by differential scanning calorimeter might improve the ECE value.

The cooling system for microsystem brings a high demand on the solid refrigeration at different temperature ranges.
Both in Fig.~\ref{fig:figure3}(a) and Fig.~\ref{fig:figure3}(b) it reveals that the ECE of BZT-12 and BZT-20 has significantly higher value than pure BTO at certain temperature range.
This might provide an application potential for the micro cooling system.
Apart from the above potential application, some other benefits of BZT, e.g., a wide range of operating temperature and environmental friendliness, are obtainable.

Apart from the investigations on BTO, BZT-12 and BZT-20 by the simulation and experiment, further simulations are done on BaZr$_{x}$Ti$_{1-x}$O$_3$ with $ 0.3 \leq x \leq 0.7 $, as shown in Fig.~\ref{fig:figure3}(c). 
Compared with the range of $ 0 \leq x \leq 0.2$, in the range of $ 0.3 \leq x \leq 0.7 $, 
the ECE is weakened gently due to the fact that $N_{effective}$ increases not so severely as in the range of $ 0.1 \leq x \leq 0.2 $.
With further increase of Zr concentration, $N_{effective}$ seems to become saturated. Therefore, the value of the ECE peak decreases slightly.
In the concentration range of $ x \leq 0.8 $, there is no ECE peak within the temperature range studied. It may be explained by the full saturation of $N_{effective}$. In addition, the temperature variation falls off with the increasing temperature. The phenomenon in pure BaZrO$_3$ agrees qualitatively with the theoretical result on SrTiO$3$ without considering misfit strain~\cite{2012_Zhang}.

\begin{figure*}[htp]
  \centering
  \centerline{\includegraphics[width=12cm]{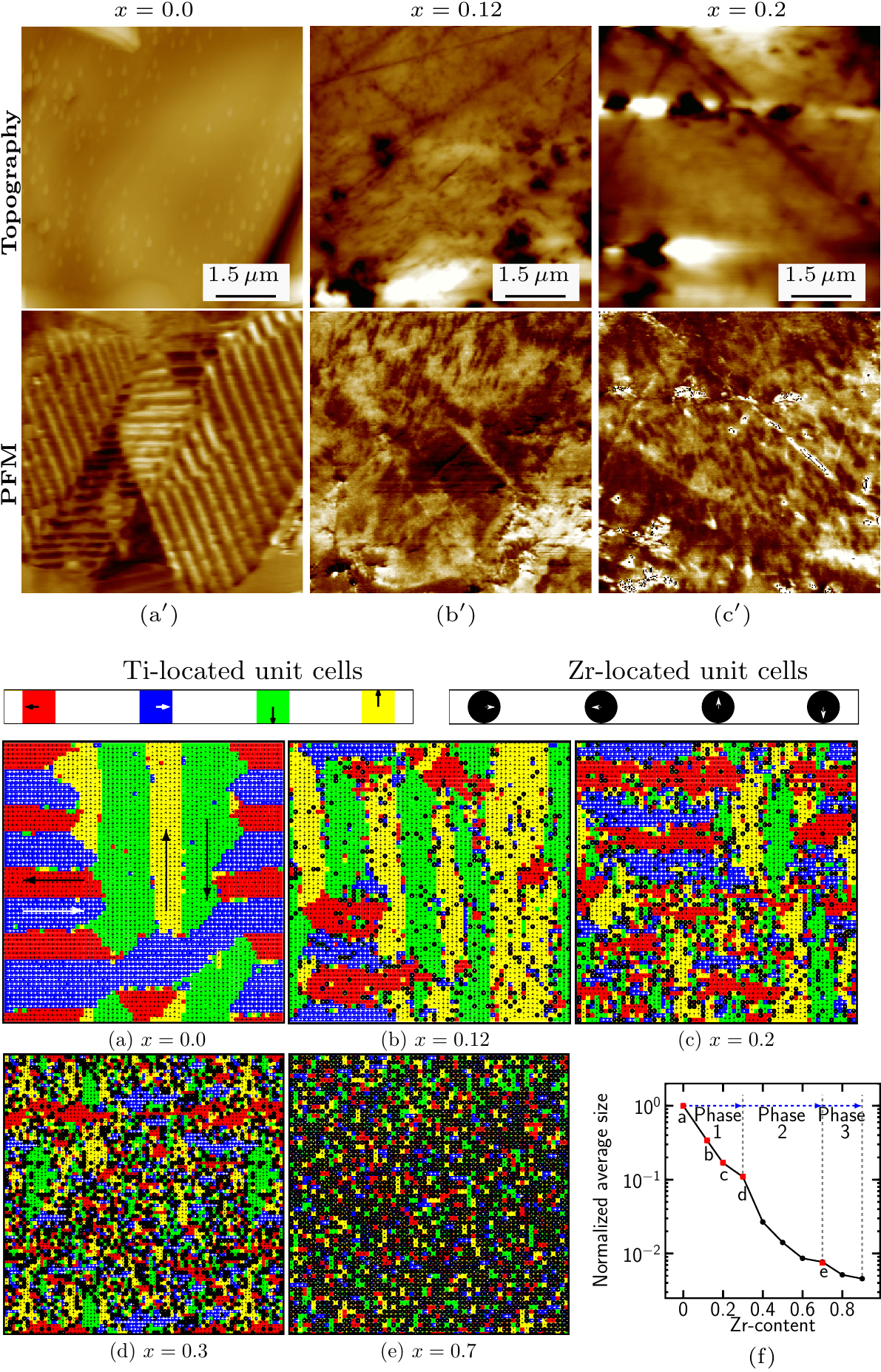}}
   \caption{
   (a$^\prime$)-(c$^\prime$) topography and Piezoresponse Force Microscopy (PFM) images , (a)-(e) the simulated domain patterns at room temperature, and (f) the normalized domain size.
   Topography and PFM images for BTO, BZT-12 and BZT-20 are shown. 
   Multiple regular ferroelectric domains are observable in BTO.
   However, in BZT-12 and BZT-20 instead of the regular domains the mosaic-like domain patterns are observed.
   With increasing Zr content, the domain size decreases, which suggests the crossover from the ferroelectric state to the relaxor state and supports the claim in the simulation.
   In the simulation, the domain patterns are shown for BaZr$_{x}$Ti$_{1-x}$O$_3$ from (a) to (e). 
   The miniaturization of the domain size is observable upon increasing Zr concentration. 
   Domain size decreases moderately with $ 0 \leq x \leq 0.2$. Sharp change of domain size appears when approaching $ x = 0.3$, which is a hint of the onset of the relaxor behavior. With $ 0.3 \leq x \leq 0.6$ almost no long-range order can be observed since the sites occupied by Zr-located unit cell become dominant. Polar clusters are observed with $0.7 \leq x \leq 0.9$, which suggests the disappearance of the relaxor behavior. No polar sites can be observed in pure BaZrO$_3$, since BaZrO$_3$ is a paraelectric material.
   According to the domain pattern, the domain size is calculated with respect to the Zr content in (f). 
  The values of domain sizes are normalized by average size of pure BTO.
  Big domain sizes are present with Zr content lower than 0.2, which is the ferroelectric state.
  With Zr content more than 0.3, the polar nanoregions appear since it is in the relaxor state.
  Furthermore, no domain patterns can be observed when Zr content is higher than 0.7 in BZT, which indicates the existence of dipolar cluster.}
 \label{fig:figure6}
\end{figure*}

\section{State transition}\label{sec:eq}
The ECE behaivor of BZT is closely related to the state transition and the polarization switching of the materials, which deserve detailed investigations as shown in this section.

\subsection{Temperature-induced polarization}
The average polarization over MC steps and space was demonstrated in Fig.~\ref{fig:figure5} with respect to the temperature. 
It can be seen that for pure BTO the average polarization changes sharply around $T=400$\,K, which corresponds to 
the transition temperature from the tetragonal to cubic phase. 
For BZT, since the sites occupied by Zr-located unit cells are weakly polar or even nonpolar, less thermal energy is required to achieve the transformation from the order to disorder phase. 
Wherefore, upon increasing Zr content in BZT, the temperature where the average polarization becomes almost zero shifts to a lower temperature. 
Meanwhile, higher Zr concentration makes the temperature-induced polarization change more moderate.
Additionally, at the same temperature the average polarization is lower when more sites occupied by Zr-located unit cells are introduced.
For better understanding of this phenomena, the corresponding domain structures at $T=300$K, are visualized in Fig.~\ref{fig:figure6} and discussed in the following subsection.


\subsection{Domain structure}

\begin{figure}[t]
  \centering
  \centerline{\includegraphics[width=7.5cm]{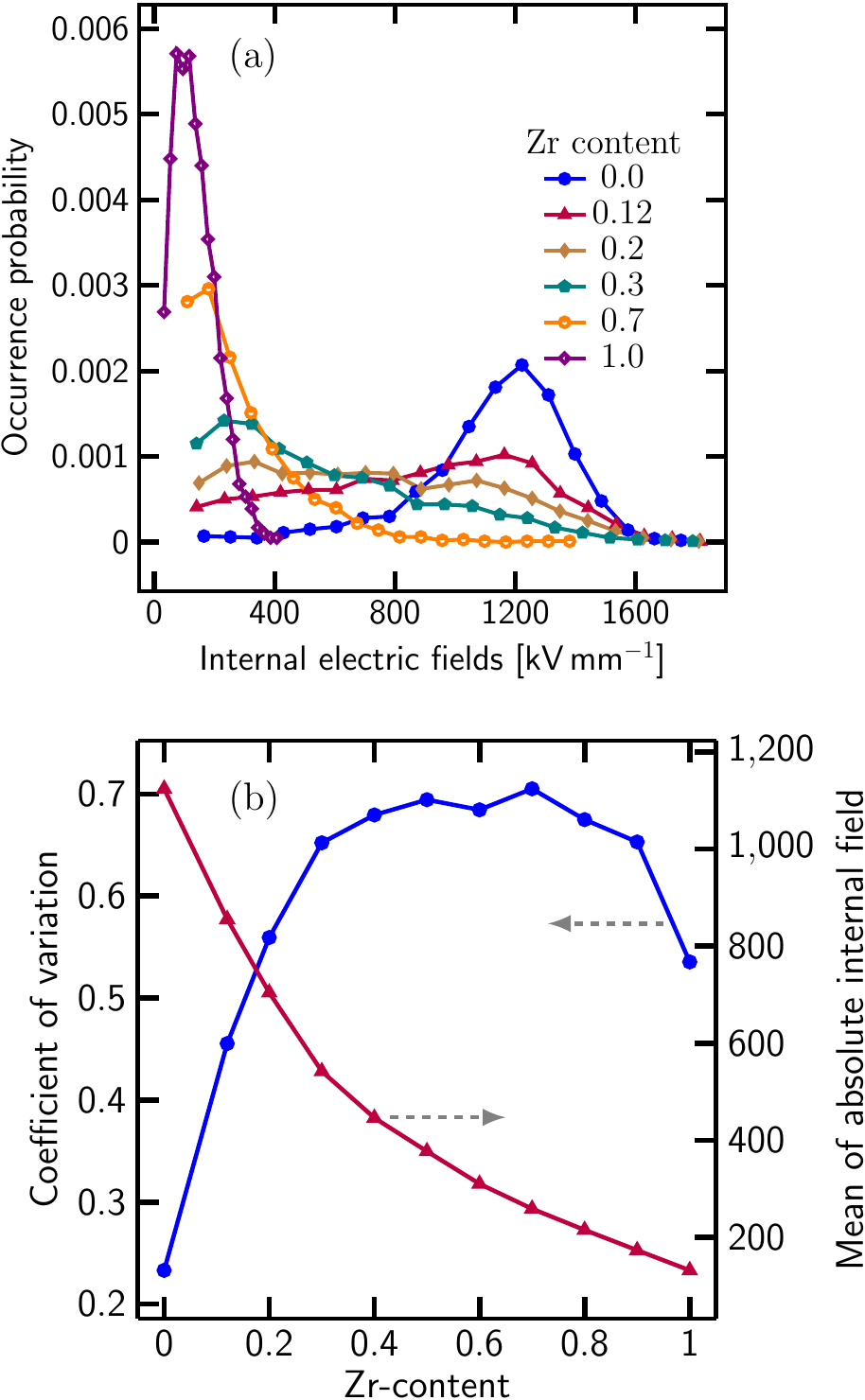}}
   \caption{
   (a) the occurrence probability distribution, and (b) the coefficient of variation of the absolute internal electric field.
   In the simulation, based on the spatial distribution of the polarizations in Fig.~\ref{fig:figure6}, the magnitude of the internal electric fields imposed on each site can be calculated. 
   (a) shows that the distribution peak shifts to lower field with more Zr substitution. 
   (b) presents that the internal fields are the internal fields are most random for BaZr$_{x}$Ti$_{1-x}$O$_3$ with $ 0.3 \leq x \leq 0.9 $. 
   The appearance of the relaxor behavior relies mainly on the coefficient of variation rather than the magnitude of the internal fields.
   Namely, with a big coefficient of variation, the relaxor behavior shows up even under a high magnitude of the internal fields.
      }
 \label{fig:figure7}
\end{figure}

In Fig.~\ref{fig:figure6} at room temperature the equilibrium domain patterns recorded investigated by PFM ((a$^\prime$)-(c$^\prime$)) and simulation ((a)-(e)). In (f), based on the simulated spatial distribution of the polarizations, the internal electric fields imposed on each site can be evaluated and analyzed.
The long-range correlated Ti-Ti sites are interrupted by the sites occupied by Zr-located unit cells~\cite{2009_Shvartsman} since Zr-site is weakly polar or nonpolar. 
Hence, in general the miniaturization of the domains is observable upon increasing Zr concentration.

In Fig.~\ref{fig:figure6}(a$^\prime$), (b$^\prime$), (c$^\prime$), (a), (b), and (c) BTO, BZT-12 and BZT-20 are investigated.
The PFM images were acquired on the polished surface within the area of one single grain.
At room temperature, BTO has a narrow distribution of large internal fields as can be seen in Fig.~\ref{fig:figure7}, which bring out a high order in the samples.
As expected the stripe-like regular domain patters are observed for BTO in Fig.~\ref{fig:figure6}(a$^\prime$).
Similar kind of the domains can be observed in the simulation (see Fig.~\ref{fig:figure6}(a)).
Both the experiment and the simulation confirm that BTO has a ferroelectric phase at room temperature.
In BZT-12 and BZT-20 the distribution of the internal fields is broad, and even no apparent peaks can be observed (see Fig.~\ref{fig:figure7}), which leads to the disorder in the samples. 
Therefore, instead of regular domains only mosaic-like domain patters exist in BZT-12 and BZT-20 (see Fig.~\ref{fig:figure6}(b$^\prime$) and (c$^\prime$)).
The domains in BZT-12 are larger than those in BZT-20. 
Similar phenomena are observed also in the simulated domain patterns (see Fig.~\ref{fig:figure6}(b) and (c)).
In brevity, the decrease of domain size with increasing the Zr content suggests the crossover from the ferroelectric state to the relaxor state.
As will be proposed in the discussions for the simulation results, random fields are responsible for these observations.
These random fields can arise from the breakdown of long-range well correlation of Ti-Ti sites due to the presence of nonpolar or weakly polar sites occupied by Zr-located unit cells~\cite{2009_Shvartsman}.
Moreover, the correlation strength between sites occupied by Ti- and Zr-located unit cells is weakened by increasing the Zr content, which leads to additional contribution to the random fields. 
The high-frequency permittivity is composition-dependent so that the correlation strength between sites can be both composition- and position-dependent if referring to Eq.~\ref{eq:dipole_inter}.

Additionally, several composition ranges can be distinguished by observing the simulated domains pattern in Fig.~\ref{fig:figure6}(a)-(e).
In the range of $ 0 \leq x \leq 0.2$ the domain size is comparably large.
The phenomenon is probably due to the persisting existence of ferroelectric since a large portion of the sites are occupied by Ti-located unit cells still and dominate the material behavior.
In the range of $ 0.3 \leq x \leq 0.6$ enough Zr-located unit cell sites are incorporated, and sufficient area can be influenced by random fields. 
Hereby random fields arise mainly from two parts: 1) the discrepancy of multi-well potential for Ti$^{4+}$ sites and single well potential for Zr$^{4+}$ sites, and 2) the random correlation length between sites due to the composition-dependent dipole-dipole interaction.
Therefore, the domain patterns are unstable and resemble as polar nanoregions when approaching $ x = 0.3$, which hints the onset of the relaxor behavior.
When Zr content is above 0.3, enough sites are occupied by Zr-located unit cells, and dominate the materials behavior. 
Therefore, no long-range order can be observed.
In the composition range of $ 0.7 \leq x \leq 0.9$, BZT can be interpreted as the nonpolar Zr matrix with introduction of the polar Ti-sites, since the Zr content is fairly high. These dipolar clusters are almost isolated from each other, and the relaxor-like behavior is not expected.
Finally, in the case of pure BaZrO$_3$ no dipolar clusters exist because BaZrO$_3$ is a paraelectric material.

According to the domain patterns shown above, the domain sizes were analyzed in Fig.~\ref{fig:figure6}(f). It should be noted that the domain size shown in Fig.~\ref{fig:figure6}(f) is normalized by the average domain size of pure BTO.
Likewise, the distinguished composition ranges mentioned above can be also identified in this plot.

\begin{figure*}[t]
  \centering
  \centerline{\includegraphics[width=17cm]{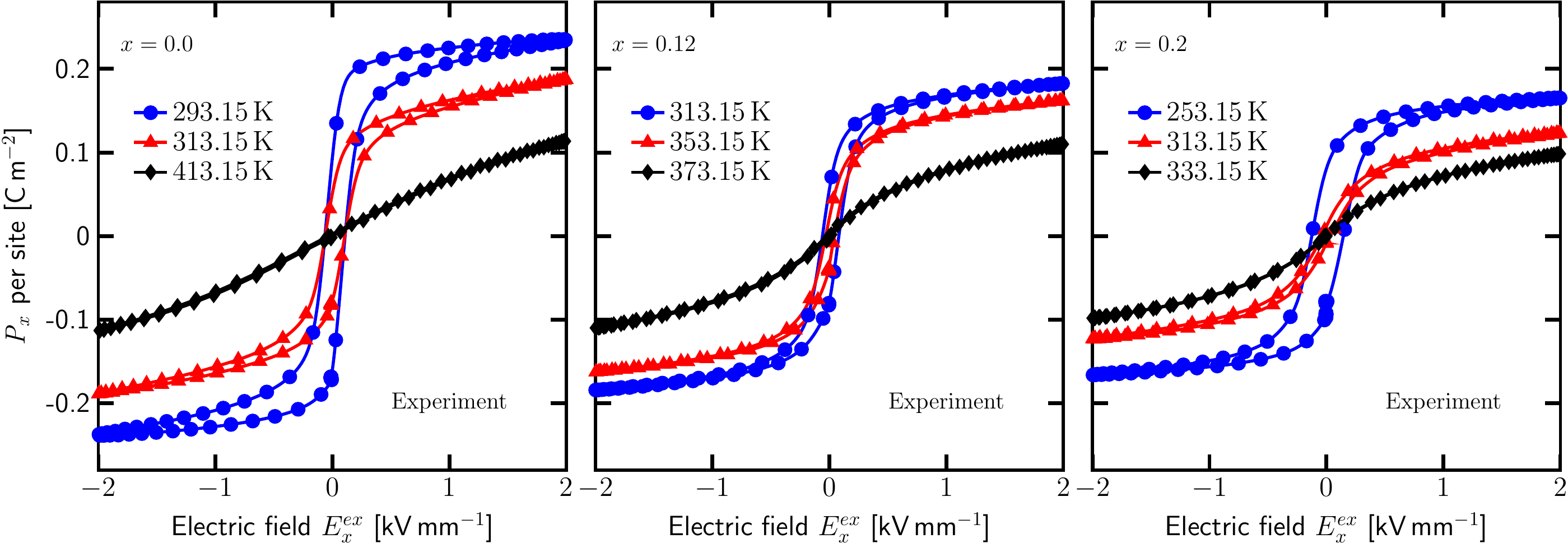}}
   \caption{The measured dielectric hysteresis at different temperatures for BaTiO$_3$, BaZr$_{0.12}$Ti$_{0.88}$O$_3$ and BZT-20. 
   At higher temperature or with increasing Zr content, the hysteresis loops become slimmer.}
 \label{fig:figure8}
\end{figure*}

\begin{figure*}[!htp]
  \centering
  \centerline{\includegraphics[width=17.5cm]{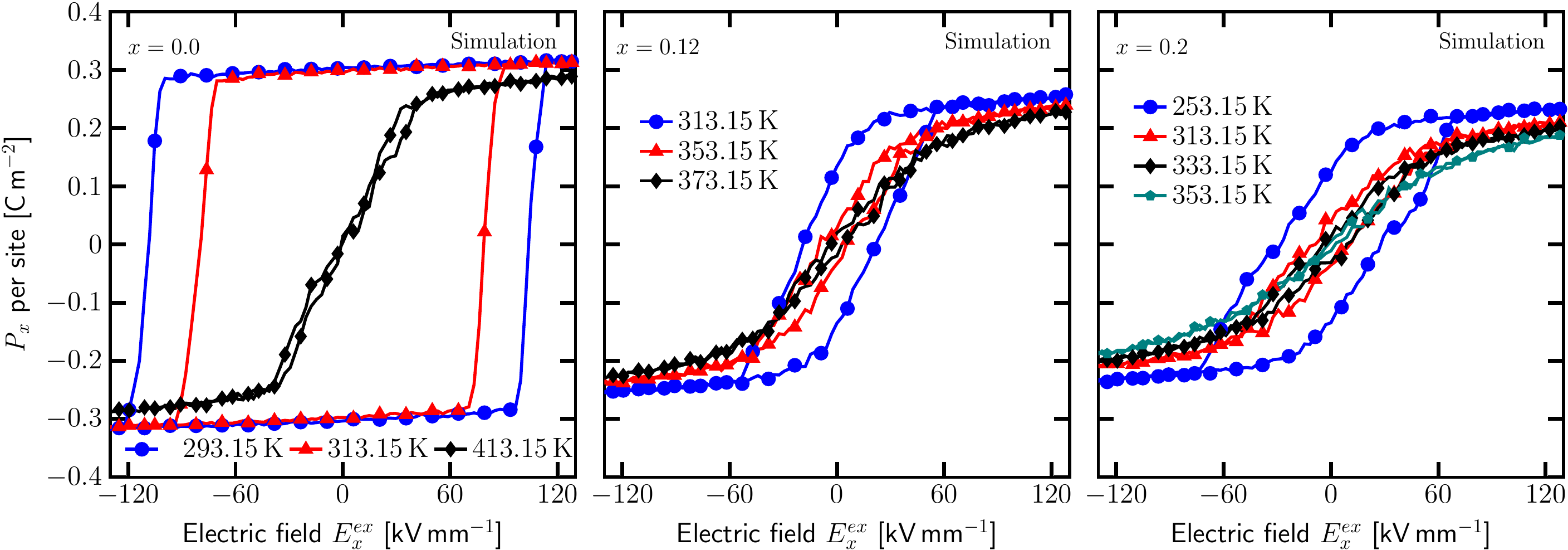}}
   \caption{The simulated dielectric hysteresis at various temperatures for BaTiO$_3$, BaZr$_{0.12}$Ti$_{0.88}$O$_3$ and BaZr$_{0.2}$Ti$_{0.8}$O$_3$. 
   It shows that either by increasing the temperature or raising Zr content, the hysteresis loops become slimmer, i.e., with lower remnant and saturation polarization and smaller coercive field. These results agree with the experimental work in Fig.~\ref{fig:figure8} qualitatively.}
 \label{fig:figure9}
\end{figure*}

\begin{figure*}[t]
  \centering
  \centerline{\includegraphics[width=12cm]{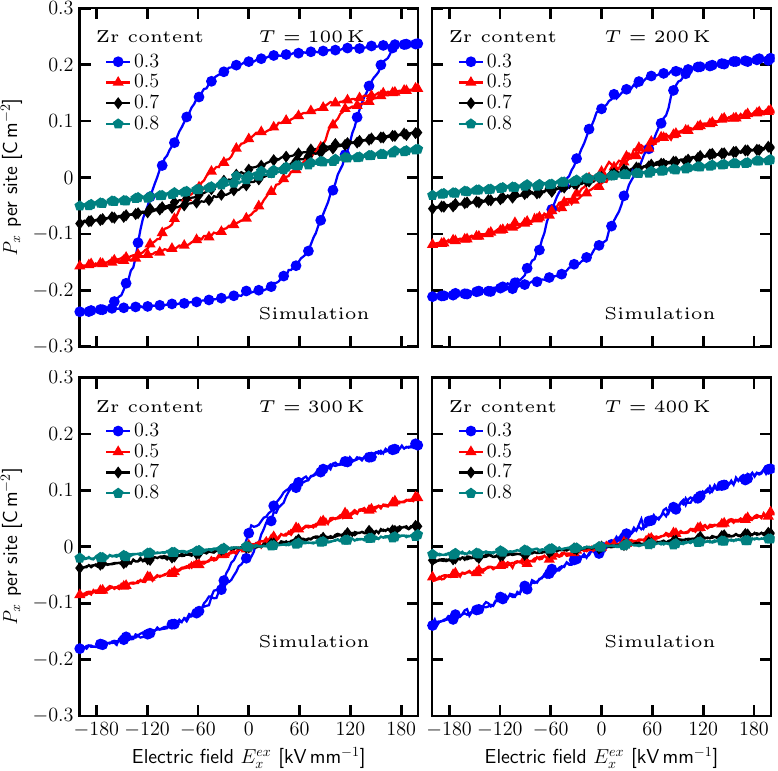}}
   \caption{The simulated dielectric hysteresis at various temperatures for BaZr$_{x}$Ti$_{1-x}$O$_3$ with $x \geq 0.3$. 
   For compositions
of $0.3 \leq x \leq 0.7 $ the relaxor-like hysteresis is present.}
 \label{fig:figure10}
\end{figure*}

%
Fig~\ref{fig:figure7}(a) shows the occurrence probability distribution of the magnitude of the internal fields for different Zr contents. It can be observed that the peak of the occurrence probability shifts to lower field upon increasing Zr content. The coefficient of variation (CV) and the mean value of the fields are also presented in Fig.~\ref{fig:figure7}(b). The CVs forms a vault-like curve during the increase of Zr concentration, while the mean value decreases gradually. 
For pure BTO, Fig.~\ref{fig:figure7}(a) shows that the occurrence probability distributes narrowly and in the high field range, which is compatible with the well small CV and a big mean value in Fig.~\ref{fig:figure7}(b).
This indicates that an apparent ferroelectric domain pattern can be formed in presence of a well distributed internal field, which keeps coherent to Fig.~\ref{fig:figure6}(a).
For BZT-12 and BZT-20, the occurrence probability distributes wider than pure BTO.
However, for BZT-12 and BZT-20 the field distributes most probably in the high field region, and the CVs are still smaller than BZT with $x \geq 0.3$.
It can be inferred that the domains undergo a miniaturization, but BZT-12 and BZT-20 still belong to ferroelectric material. 
Fig.~\ref{fig:figure6}(b) and (c) substantiates this inference.
For $ 0.3 \leq x \leq 0.6$ in BZT, CVs possess a large value while the mean value of fields is moderate.
It can be concluded that the internal fields are large and random, which dominate the materials behavior.
In other words, it leads to the relaxor states with polar nanoregions (see Fig.~\ref{fig:figure6}(d)).
For BZT-80 and BZT-90, CVs are still large, and simultaneously the mean value of fields are quite small.
It suggests that the fields are small and random, which corresponds to the appearance of dipolar clusters (see Fig.~\ref{fig:figure6}(e)).
For pure BaZrO$_3$, the CVs are sharply drops to a small value with a small mean value of fields.
It stands for a state with small and narrow distributed fields, which results in the paraelectric phase, and no domains structures can be observed.


\subsection{Hysteresis}

The P-E loops of BZT with Zr content $x=0.0$, $x=0.12$ and $x=0.2$ are investigated both experimentally and theoretically at different temperatures, as can be seen in Fig.~\ref{fig:figure8} and Fig.~\ref{fig:figure9}.

In BTO the tetragonal phase transforms to the orthorhombic phase around 293.15\,K, which is indicated by an ECE peak in Fig.~\ref{fig:figure3}(a). 
Therefore, there is a sudden drop of the saturation polarization $P_\mathrm{sa}$ and the remnant polarization $P_\mathrm{r}$ from 293.15\,K to 313.15\,K (see Fig.~\ref{fig:figure8}).
In the simulation, only the transition from the cubic to tetragonal phase is considered. Hence, this sudden drop cannot be observed (see Fig.~\ref{fig:figure9}). 
Above 313.15\,K the results for BTO in Fig.~\ref{fig:figure8} agree with the simulation results qualitatively (see Fig.~\ref{fig:figure9}). 
Similarly, the sharp variations of $P_\mathrm{sa}$ and $P_\mathrm{r}$ in BTO from 393.15\,K to 413.15\,K, in BZT-12 from 353.15\,K to 373.15\,K and in BZT-20 from 313.15\,K to 333.15\,K indicate the transition from the tetragonal to cubic phase, which is responsible for the ECE peaks in Fig.~\ref{fig:figure3}(a).

Under sinusoidal external electric field, the polarization switching process can be studied by canonical MC simulations (see Fig.~\ref{fig:figure9}).
It can be seen that the hysteresis loops become smaller when the temperature or the Zr concentration increases, more exactly, $P_\mathrm{sa}$, $P_\mathrm{r}$ and the coercive field $E_c$ decrease. 
As temperature increases, higher thermal energy promotes the thermal fluctuation of the polarization, which eliminates the necessity of high electric field to achieve the domain switching.
Experimental data~\cite{2002_Yu, 2014_Mahesh} show decrease of $P_\mathrm{sa}$, $P_\mathrm{r}$ and $E_c$ with increasing temperature, which agrees with our simulation qualitatively.
As for the influence of Zr concentration, it can be reasoned that at weakly polar or nonpolar sites occupied by Zr-located unit cells it is more effortless to reverse the polarization, compared with polar sites occupied by Ti-located unit cells.

The variation of the loops with respect to the Zr concentration can be discussed in more details.
For pure BTO or BZT with low Zr concentration ($x<0.20$), the materials are still grouped into ferroelectrics. 
Hence, the polarization switching is quite steep around $E_c$ below the phase transition temperature.
In the experimental work~\cite{2008_Moura}, the well saturated hysteresis loops with regular shape for BZT-5, BZT-10 and BZT-15 were observed, which is typical for ferroelectric materials and supports our conclusion. 
Even by introducing only a small amount of Zr ($x=0.12$) the coercive field is sharply reduced inasmuch as the sites occupied by Zr-located unit cells can be the nucleation source for reversed domains.
By contrast, a moderate decrease of coercive field is observed upon further increasing Zr content.

In short, both in the measurement and the simulation at higher temperature or with increasing Zr content, the hysteresis loops become smaller, and the slope around the coercive field becomes less steeper.
The saturation polarization in the simulation is higher than that in the experiment, since it is typical that the single crystal without defects has a higher saturation polarization than the polycrystalline ceramics.

Additionally, more compositions are investigated to reveal the influence of Zr content on the materials behavior (see Fig.~\ref{fig:figure10} ).
A relaxor-like hysteresis can be seen for compositions of $0.3 \leq x \leq 0.7 $, even at $T=100$\,K. In the experiments by Yu \textit{et al.}~\cite{2002_Yu}, the relaxor-like hysteresis was also observed in BZT-30 at quite low temperature $T=175$\,K.
At $T=100$\,K with $x \leq 0.7 $, $P_\mathrm{r}$ is still nonzero while $P_\mathrm{r}$ is almost zero with $x \geq 0.8 $. 
This phenomenon keeps in accordance with the observation in Fig~\ref{fig:figure5}, i.e., the average polarization becomes almost zero only when $x \geq 0.8$ at $T=100$\,K.
The same explanation can be extended to interpret the nil value of $P_\mathrm{r}$ at $T=200$\,K with $x \geq 0.6$ and at $T=300$\,K with $x \geq 0.4$.
In BZT with $x=0.8$ and $x=0.9$, only dipolar clusters exist. 
Merely at quite low temperatures (e.g., $T=100$\,K), these dipolar clusters can be correlated. 
Therefore, the hysteresis can be solely observed at low temperature.
At a higher temperature (e.g., $T=200$\,K), the weak mutual interaction of these dipolar clusters is overwhelmed by a high thermal fluctuation. 
Hence, a linear type of dielectric response is expected.
Furthermore, the hysteresis loops nearly overlap with each other for these two compositions.
For pure BaZrO$_3$, i.e., paraelectric phase, the
dielectric hysteresis persistently shows a line type within the investigated temperature range.

\section{Conclusions}
Through combination of the canonical MC and microcanonical MC, we propose a lattice-based MC scheme to evaluate directly the ECE in BaZr$_{x}$Ti$_{1-x}$O$_3$. Within this scheme, the sites occupied by Ti-located unit cells are described by a multi-well type of Ginzburg-Landau term while the sites occupied by Zr-located unit cells by a single-well type. Simultaneously, the high-frequency permittivity, which reflects the inverse of the dipole-dipole interaction strength, is assumed to be a composition-dependent parameter.

Firstly, the ECE measurement and the simulation are carried out for BTO, BZT-12 and BZT-20. Both the experiment and simulation reveal that a sharp drop of ECE peak was observed with increasing the Zr substitution from 0\% to 20\%.
Meanwhile, with higher Zr content, the ECE peak shifts to lower temperatures.
Regarding to the peak of the ECE three phases can be distinguished in the simulation: a sharp drop of the peak with $ 0.0 \leq x \leq 0.2 $, a moderate drop with $ 0.3 \leq x \leq 0.7 $, a very weak peak or no peak existence with $x \geq 0.8 $.

Whereafter, the composition-dependence of the domain patterns, and that of the hysteresis was investigated in order to understand two important influential factors of the ECE: the state transition and the polarization switching.
First of all, the domain patterns at room temperature were acquired through PFM. In BTO, a regular strip-like ferroelectric domains are present. 
However, instead of regular domain pattern, a mosaic-like domain patterns are present in BZT-12 and BZT-20. In BZT-12 the domain size is bigger than in BZT-20.
All these facts indicate a crossover from the ferroelectrics to the relaxor ferroelectrics, which agrees with the simulation. 
Moreover, the simulated domain patterns can visualize several different phases, including the phase with big domain size, the phase with small domain size and dipolar clusters respectively.
According to the above domain patterns the internal electric fields were calculated, and the probability distribution of the fields is illustrated and analyzed. 
Especially, the vault-like coefficient of the variation of fields reveals several transition step by step with increasing Zr content in BZT: ferroelectrics, relaxor ferroelectrics, dipolar clusters and paraelectrics.
The polarization switching behavior was additionally investigated in the hysteresis experimentally and theoretically. 
The hysteresis reveals that upon increasing Zr concentration or the temperature, the remnant and saturation polarization and the coercive field decrease.
Meanwhile, the simulations reveals that with $x\geq 0.3$ the hysteresis shows a typical relaxor-type.


This work reveals that at certain temperature, the ECE of BZT-12 or BZT-20 is more significant than BTO. This provides an application potential of BaZr$_{x}$Ti$_{1-x}$O$_3$ for the cooling devices. 
In BaZr$_{x}$Ti$_{1-x}$O$_3$ another advantage disclosed by this work is a wider application temperature range than in BTO, which is essential for the application.
Also, in this work we intensively investigated the state transition, which might benefit other researchers to understand the relaxor behavior.

\appendix

\section{Algorithms} \label{sec:alg}

In the canonical ensemble, the Metropolis Monte-Carlo method is utilized to simulate the materials behavior at constant temperature. 

For the evaluation of ECE there are the so-called indirect method by using Maxwell relations~\cite{2009_Caoa,2011_Pirc,2011_Dunne} and the direct method by using Creutz's algorithm~\cite{2012_Ponomareva}. In the direct method, the thermal energy is presented as:
\begin{equation}
E_k = \sum_{i,k} Demon(\mathbf{r}_i,k) = \frac{f}{2   }N k_BT,
\label{eq:thermal}
\end{equation}
where $Demon(\mathbf{r}_i,k)$ represents the thermal energy of the two degrees of freedom carried by the $k$-th demon at site $i$, $f$ is the number of degrees of freedom per site, and $N$ is the total number of sites. 
In microcanonical ensemble with constant energy, if $\Delta E(\mathbf{r}_i)<0$ or $Demon(\mathbf{r}_j,k)>\Delta E(\mathbf{r}_i)$ the switching is approved, and the demon energy is updated: $Demon(\mathbf{r}_j,k)=Demon(\mathbf{r}_j,k)-\Delta E(\mathbf{r}_i)$. 

The total thermal energy of a system is given as in
Eq.~(\ref{eq:thermal}).
Since in BZT there are 5 atoms per lattice site, i.e., $10$ degrees of freedom in 2D, 5 demons per site are considered in this paper. It should be noted that since in 3D case the number of degrees of freedom per site is 15, the heat capacity in our simulation is underestimated, leading to the overestimation of the temperature variation. The periodic boundary condition is utilized so that the surface effect can be ignored.


\begin{acknowledgments}
The funding of Deutsche Forschungsgemeinschaft
(DFG) through project B3, B5 and B4 (XU 121/1-2, AL 578/16-2, GE 2078/3-2, LU
729/15-2) within the
Priority Programme Caloric Effects in Ferroic Materials:
New Concepts for Cooling (SPP 1599) is gratefully acknowledged.
Competence Center of High Performance Computing in Hesse (HPC Hessen) is appreciated for the calculation resources.
Thanks for the useful discussions from Prof. George Rossetti in University of Connecticut; Dr. Yuri Genenko, Kai-Christian Meyer in TU Darmstadt.
\end{acknowledgments}

%

%

\end{document}